\documentclass[12pt]{iopart}

\usepackage{iopams}
\usepackage{graphicx}
\begin{document}

\title[]{Fermi Surface Variation of Ce 4$f$-electrons in Hybridization Controlled Heavy-Fermion Systems}

\author{H. J. Im$^{1,2,3}$, T. Ito$^{2,4,5}$, H. Miyazaki$^{2,5,6}$, S. Kimura$^{2,4,7}$, Y. S. Kwon$^{8}$, Y. Saitoh$^9$, S.-I. Fujimori$^9$, A. Yasui$^{9,10}$ and H. Yamagami$^{9,11}$}
\address{$^1$ Department of Physics, Sungkyunkwan University, Suwon 440-746, Republic of Korea}
\address{$^2$ UVSOR Facility, Institute for Molecular Science, Okazaki 444-8585, Japan}
\address{$^3$ Department of Advanced Physics, Hirosaki University, Hirosaki 036-8561, Japan}
\address{$^4$ School of Physical Sciences, The Graduate University for Advanced Studies, Okazaki 444-8585, Japan}
\address{$^5$ Graduate School of Engineering, Nagoya University, Nagoya 464-8603, Japan}
\address{$^6$ Center for Fostering Young and Innovative Researchers, Nagoya Institute of Technology, Nagoya 466-8555, Japan}
\address{$^7$ Graduate School of Frontier Biosciences, Osaka University, Suita 565-0871, Japan}
\address{$^8$ Department of Emerging Materials Science, Daegu Gyeongbuk Institute of Science and Technology (DGIST), Daegu, 711-873, Republic of Korea}
\address{$^9$ Synchrotron Radiation Research Center, Japan Atomic Energy Agency, SPring-8, Sayo, Hyogo 679-5148, Japan}
\address{$^{10}$ Japan Synchrotron Radiation Research Institute/SPring-8, Sayo, Hyogo 679-5198, Japan}
\address{$^{11}$ Department of Physics, Kyoto Sangyo University, Kyoto 603-8555, Japan}
\ead{hojun@cc.hirosaki-u.ac.jp}


\begin{abstract}
Ce 3$d$-4$f$ resonant angle-resolved photoemission measurements on CeCoGe$_{1.2}$Si$_{0.8}$ and CeCoSi$_{2}$ have been performed to understand the Fermi surface topology as a function of hybridization strength between Ce 4$f$- and conduction electrons in heavy-fermion systems.
We directly observe that the hole-like Ce 4$f$-Fermi surfaces of CeCoSi$_{2}$ is smaller than that of CeCoGe$_{1.2}$Si$_{0.8}$, indicating the evolution of the Ce 4$f$-Fermi surface with the increase of the hybridization strength.
In comparision with LDA calculation, the Fermi surface variation cannot be understood even though the overall electronic structure are roughly explained, indicating the importance of strong correlation effects.
We also discuss the relation between the Ce 4$f$-Fermi surface variation and the Kondo peaks.

\end{abstract}

\maketitle

\section{Introduction}
In Ce-based metals, the Fermi surface (FS) topology of strongly correlated Ce 4$f$-electrons directly influences unusual physical properties such as heavy-fermion behavior and quantum criticality \cite{Stew84,Gege08}.
It has been revealed in many experimental and theoretical studies that strongly correlated Ce 4$f$-electrons form large density of states at the Fermi level ($E_F$) through the hybridization with conduction electrons ($c$-$f$ hybridization).
Such large density of states is called Kondo resonance peak and is the origin of a large effective mass of charge carriers in heavy-fermion systems \cite{Allen86,Malt96,Patt90,Gunn83}.
In addition, as a function of the hybridization strength, their ground state changes from magnetic to non-magnetic heavy-fermion via a quantum critical point (QCP) \cite{Doni77}.
Recently, the variation of Ce 4$f$-FS through QCP has been considered as a crucial phenomenon to distinguish two contrast scenarios for the quantum criticality:
One is a spin-density wave (SDW) scenario \cite{Mill93} where the Ce 4$f$-FS changes continuously through QCP as in CeRu$_2$Si$_2$ \cite{Daou06} and CeIn$_3$ \cite{Kuch06}.
The other is a local quantum critical senario \cite{Si01} where the Ce 4$f$-FS changes discretely through QCP as in CeRhIn$_5$ \cite{Shis05} and CeCu$_{0.9}$Au$_{0.1}$ \cite{Schr00}.
Therefore, it is essential to understand how the FS of Ce 4$f$ electrons forms and changes as a function of hybridization strength.
To this end, the angle-resolved photoemission spectroscopy (ARPES) is one of the most promising experimental method because the FS can be directly observed together with the band dispersion which provides a fruitful information of the electronic structure \cite{Hufn95}.
By combining angle-resolved and resonance photoemission spectroscopies on CeCoGe$_{1.2}$Si$_{0.8}$, we have clearly observed the dispersive Kondo resonance peaks crossing $E_F$ in unoccupied regime, which form Ce 4$f$-FS though the $c$-$f$ hybridization \cite{Im08}.
This gives a good opportunity to study the Ce 4$f$-FS topology as a function of hybridization strength.
In this article, we study Ce 4$f$-FS topologies by bulk-sensitive Ce 3$d$-4$f$ resonant ARPES studies of CeCoGe$_{1.2}$Si$_{0.8}$ and CeCoSi$_{2}$, which are isostructural single crystals and have different hybridization strength.

\section{Experimetal methods}
Single crystalline samples, CeCoGe$_{1.2}$Si$_{0.8}$ and CeCoSi$_2$, crystallized in the isostructural orthorhombic CeNiSi$_2$-type (\textit{Cmcm}) structure.
They were prepared by Bridgeman and Czochralski methods with annealing at 900 $^{\circ}$C for about one week.
Laue pattern and X-ray diffraction analysis confirm that the samples used in this study have a good crystallization and are in single phase, respectively.
Samples are categorized into non-magnetic heavy-fermions and have different hybridization strength;
CeCoSi$_2$ has a larger Kondo temperature ($T_\textrm{K}$, an energy scale of c-f hybridization strength) rather than CeCoGe$_{1.2}$Si$_{0.8}$ ($T_\textrm{K}$ $\sim$ 350 K \cite{Im08}).
In these systems, the increase of hybridization strength is considered to be exclusively derived from the increase of chemical pressure as in CeNiGe$_{2-x}$Si$_{x}$ systems, where the lattice constants are reduced without the change of the Ni 3$d$ state character near $E_F$ \cite{Im07}.
This provides an ideal condition to directly compare the change of FSs caused by hybridization between the Ce 4$f$- and conduction bands.
Ce 3$d$-4$f$ resonant ARPES measurements have been performed at the BL23SU of SPring-8.
The photon energies for on- and off-resonant ARPES are set to be 886 and 879 eV from X-ray absorption spectra.
Owing to the process of the resonant PES, the on-resonant spectra mainly come from the Ce 4$f$-electronic structure and the off-resonant spectra represent the electronic structure of non-4f-states \cite{Hufn95}.
Measurement temperature and total energy resolution are about 20 K and 120 meV, respectively.
The clean surfaces in the (010) plane were prepared by \textit{in situ} cleaving of single crystal samples under a vacuum of 2 $\times$ 10$^{-8}$ Pa.
Sample cleanliness was checked by the absence of the O 1$s$ core-level spectrum.
The $E_F$ of the sample was referred to that of Au film and was calibrated by using Au 4$f$ core-level peak.

\section{Results and Discussion}

\begin{figure}
\begin{center}
\includegraphics[width=110mm,clip]{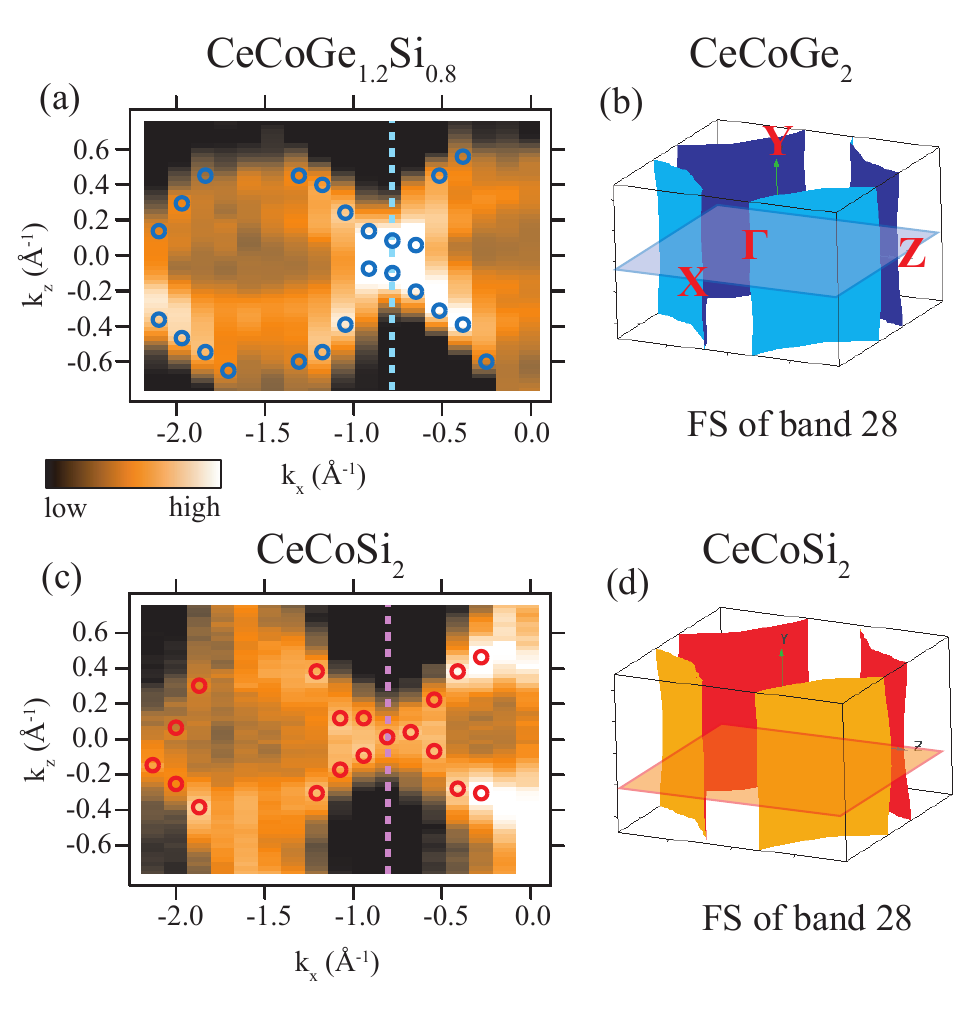}
\caption{\label{fig:figure1}
(a),(c) Ce 4$f$-FSs of CeCoGe$_{1.2}$Si$_{0.8}$ and CeCoSi$_{2}$ in the $k_z$-$k_x$ plane obtained from Ce 3$d$-4$f$ on-resonant ARPES spectra.
The open circles are obtained from the minimum of second derivative of momentum distribution curves (MDCs) at $E_F$ as shown in Fig. 2(e) and 2(f), and correspond to diamond shaped Ce 4$f$-FS contours.
(b),(d) The FSs of band 28 in the local-density approximation (LDA) band calculation of CeCoGe$_{2}$, which is a parent compound of CeCoGe$_{1.2}$Si$_{0.8}$ and CeCoSi$_{2}$.
Shade areas depict the measured momentum planes estimated from the equation, $k_{\perp} = (2m/\hbar^2(E_{kin}cos^2\theta+V_{0}))^{1/2} - k_{\perp photon}$, where $m$ is the electron mass, $E_{kin}$ the kinetic energy of the photoelectron, $V_0$ the inner potential (V$_0$ = 15.8 eV), $k_{\perp}$ the emission angle of the photoelectron relative to the surface normal and $k_{\perp photon}$ the momentum of the incident photon perpendicular to the surface ($k_{\perp photon} \approx 0.32 \textrm{ \AA}^{-1}$ for the photon energy of 886 eV) \cite{Hufn95}.
The box represents the first Brillouin zone, where $\Gamma \textrm{X} \not \approx \Gamma \textrm{Z} \approx 0.76 \textrm{ \AA}^{-1}$ and $\Gamma \textrm{Y} \approx 0.38 \textrm{ \AA}^{-1}$.}
\end{center}
\end{figure}

Figures 1(a) and 1(c) are the intensity maps in the $k_x$-$k_z$ plane, which are obtained by integrating Ce 3$d$-4$f$ on-resonant ARPES spectra from -0.1 to 0.1 eV, and represent Ce 4$f$-FSs of CeCoGe$_{1.2}$Si$_{0.8}$ and CeCoSi$_{2}$, respectively.
Measured momentum planes are depicted by the shaded planes in Brillouin zone (BZ) in Figs. 1(b) and 1(d) (for details, see the figure caption).
In both samples, we can observe the two types of FSs, i. e. diamond-shaped FSs outside and unclear FSs inside.
The diamond-shaped outside FSs correspond to the FS of band 28 which have the strong two-dimensional electronic structure as shown in Figs. 1(b) and 1(d).
These have been also observed in Ce 4$d$-4$f$ resonant ARPES measurements on CeCoGe$_{1.2}$Si$_{0.8}$ \cite{Im08}.
On the other hand, we can ascribe the inside FSs to those of band 26 and 27 which show weak two-dimensionality in LDA calculation (not shown here), even though they are too ambiguous to compare FSs between CeCoGe$_{1.2}$Si$_{0.8}$ and CeCoSi$_{2}$.
It seems that LDA calculation roughly explain the overall FS topology of both CeCoGe$_{1.2}$Si$_{0.8}$ and CeCoSi$_{2}$ as in many heavy-fermion systems \cite{Zwic92,Denl01}.
But, to be precise, it is found that the size of the outside FSs between two samples is different in ARPES while they are almost the same in LDA calculation as shown in Figs. 1(b), 1(d), and 3(b).
In order to understand the origin of the Ce 4$f$-FS variation, we here focus on the outside FSs which have strong two-dimensionality in diamond-shape and are clearly observed to compare two samples.

\begin{figure}
\begin{center}
\includegraphics[width=110mm,clip]{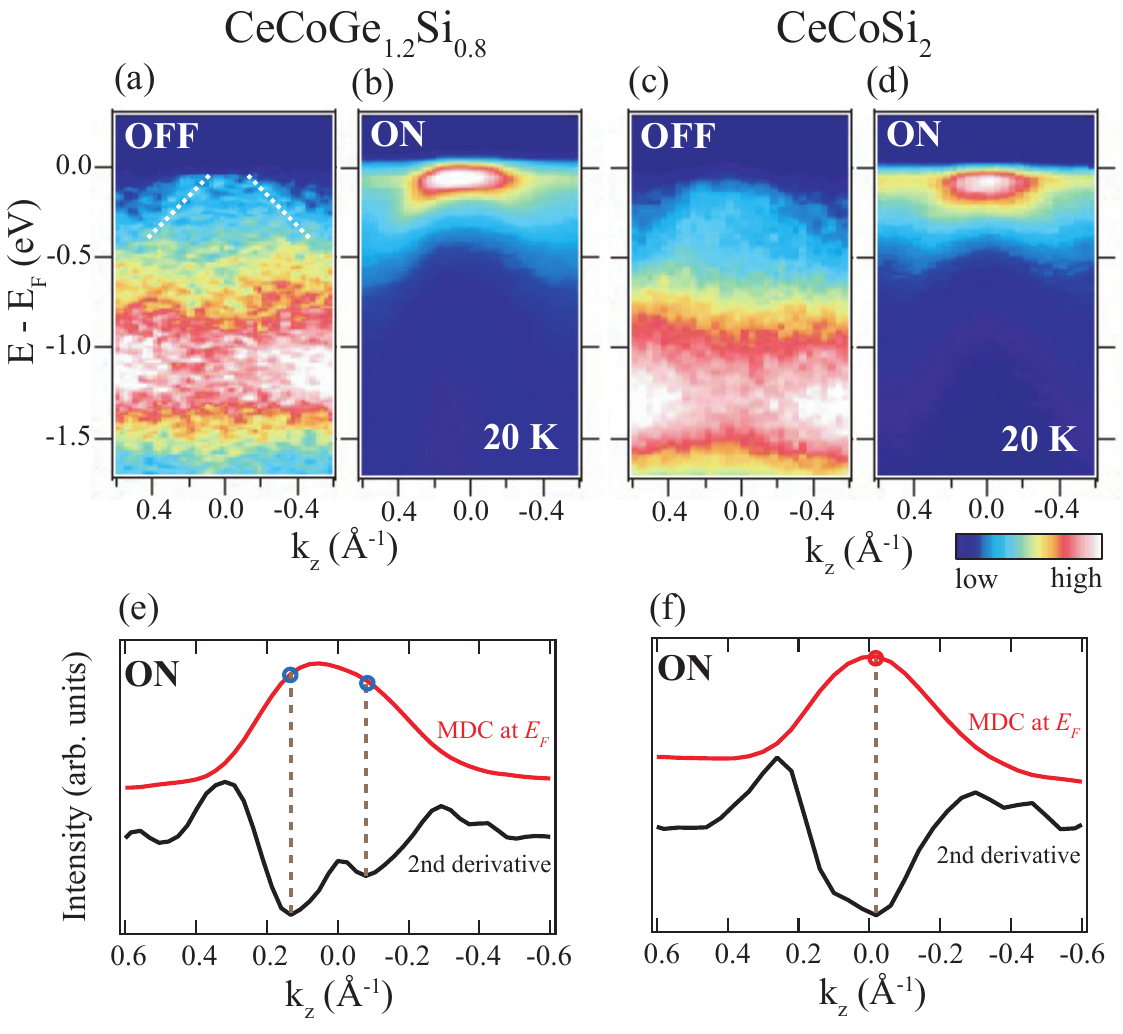}
\caption{\label{fig:figure2}
Intensity plots of Ce 3$d$-4$f$ off- and on-resonant ARPES spectra represent the band dispersion of conduction and f-electrons, respectively, along the indicated the dashed-line shown in Figs. 1(a) and 1(c) [(a),(b) for CeCoGe$_{1.2}$Si$_{0.8}$; (c),(d) for CeCoSi$_{2}$].
The dashed lines are a guide to the eye for conduction bands.
MDCs at $E_F$ of on-resonance and their second derivatives are shown for (e) CeCoGe$_{1.2}$Si$_{0.8}$ and (f) CeCoSi$_{2}$. The extreme values of the second derivative stand for $k_F$s.
}
\end{center}
\end{figure}

Figures 2(a) and 2(b) [2(c) and 2(d)] are intensity plots of the off- and on-resonant ARPES spectra of CeCoGe$_{1.2}$Si$_{0.8}$ [CeCoSi$_{2}$] along a dashed line assigned in Fig. 1(a) and 1(c), respectively:
remind that the off-spectra represent the band dispersion of conduction electrons and the on-spectra mainly do that of Ce 4$f$-electrons.
In Ce 3$d$-4$f$ off-resonant ARPES spectra, we observe the nearly flat band around -1.2 eV and the steep band near $E_F$ even though feature is not clear.
The former corresponds to mainly Co 3$d$ band as shown in AIPES study of CeCoGe$_{2}$ and band calculation \cite{Im05,Yasui09}.
The latter is the band composing the outside diamond-shaped FS in agreement with Ce 4$d$-4$f$ off-resonant ARPES results of CeCoGe$_{1.2}$Si$_{0.8}$ \cite{Im08}.
In Ce 3$d$-4$f$ on-resonant ARPES spectra, the large enhancement of the spectral weight of Ce 4$f$-bands near $E_F$ are clearly observed where the conduction bands cross $E_F$.
Such electronic structure comes from hybridization between a renormalized Ce 4$f$-band just above $E_F$ and conduction bands as in a periodic Anderson model (PAM) \cite{Im08}.
Figures 2(e) and 2(f) show the momentum distribution curves (MDCs) at $E_F$ and the second derivative of them.
We observe that the peak width of CeCoGe$_{1.2}$Si$_{0.8}$ is larger than that of CeCoSi$_{2}$ and seems not to be a single peak.
In general, the extreme values of the second derivative can be regarded as the Fermi vector ($k_F$) where a band dispersion crosses $E_F$ \cite{Yosh05}.
We obtain the double extreme values of MCD at $k_{x(z)} \sim \pm 0.15 \textrm{ \AA}^{-1}$ in CeCoGe$_{1.2}$Si$_{0.8}$, while there is only single extreme value at $k_{x(z)} \sim 0 \textrm{ \AA}^{-1}$ in CeCoSi$_{2}$.
In comparison between the FS topologies [Figs. 1(a) and 1(c)], we recognize that the one $k_F$ of CeCoSi$_{2}$ can be ascribed to the superposition of the two $k_F$s of CeCoGe$_{1.2}$Si$_{0.8}$.

Figure 3(a) is the plot of the superimposed FS contours of CeCoGe$_{1.2}$Si$_{0.8}$ and CeCoSi$_{2}$.
It is clearly observed that the hole-like Ce 4$f$-FS of CeCoGe$_{1.2}$Si$_{0.8}$ is larger than that of CeCoSi$_{2}$.
This reveals that the hole-like Ce 4$f$-FS shrinks with increasing the hybridization strength; in other words, the Ce 4$f$-electrons, which participate in the FSs, increase with the hybridization strength.
On the other hand, in LDA band calculations, the FS of CeCoGe$_{2}$ is a little smaller than that of CeCoSi$_{2}$ with almost the same size, exhibiting an opposite tendency to the results of ARPES.
This indicates that the LDA band calculation is not sufficient to explain the observed Ce 4$f$-FS variation in the different hybridization strength and that the strong correlation effects of Ce 4$f$-electrons should be considered properly\cite{Zwic92,Shim07}.

\begin{figure}
\begin{center}
\includegraphics[width=110mm,clip]{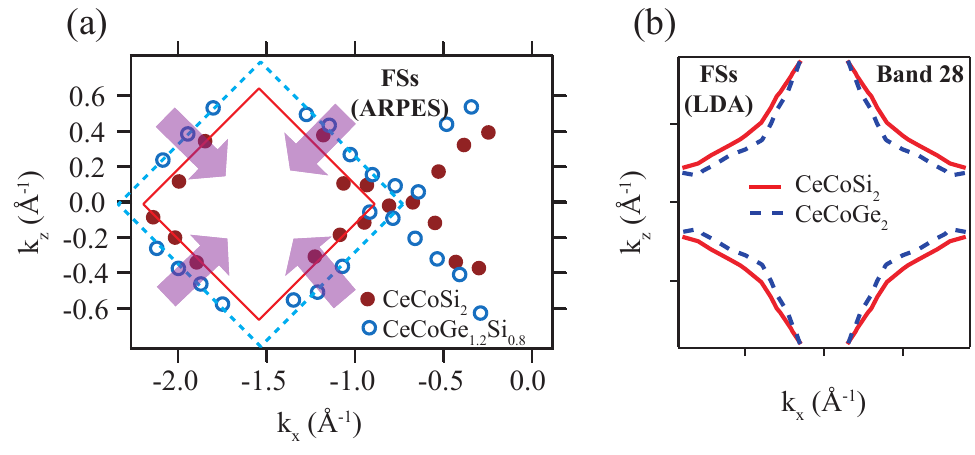}
\caption{\label{fig:figure3}
(a) Comparison of Ce 4$f$-Fermi surfaces between CeCoGe$_{1.2}$Si$_{0.8}$ (open circles) and CeCoSi$_{2}$ (solid circles).
Raw FS-contours in Figs. 1(a) and 1(c) were a little rotated to match the k-axis.
Dashed- and solid-lines are guides for eyes.
(b) In LDA calculation, the FS contours of band 28 on the measured $k$-planes in ARPES (see Figs. 1(b) and 1(d)) are almost same for CeCoGe$_{1.2}$Si$_{0.8}$ and CeCoSi$_{2}$.}
\end{center}
\end{figure}

In heavy-fermion systems, the hybridization of the strongly correlated Ce 4$f$-electrons with conduction electrons causes the separation of the spectral weight of Ce 4$f$-electrons into the coherent peak (Ce 4$f^1$; Kondo peak) around $E_F$ and the incoherent peak (Ce 4$f^0$) around -2 eV \cite{Patt90,Gunn83}.
In addition, it has been reported in many photoemission experiments that the spectral weight of $f^0$ peak around -2 eV is transferred to the weight of $f^1$ peak near $E_\textrm{F}$ with increasing the hybridization strength \cite{Malt96,Patt90,Im07,Im05}.
When we take it into account that both the Ce 4$f$ spectral weight transfer and the observed FS variation, it is suggestive that the portion of Ce 4$f$-electrons at $E_F$ increases with the hybridization strength through the spectral weight transfer and eventually the size of Ce 4$f$-FS changes.

For future study, the FS topologies of conduction electrons (non-Ce 4$f$-states) should be investigated to comprehensively understand the whole electronic structure.
It is also meaningful that the Ce 4$f$-FS variation are reproduced by theoretical approach such as a dynamical mean-field theory combining with LDA (LDA+DMFT) \cite{Shim07,Choi13}.

\section{Summary}
We have performed Ce 3$d$-4$f$ resonant ARPES experiments on CeCoGe$_{1.2}$Si$_{0.8}$ and CeCoSi$_{2}$ where the $c$-$f$ hybridization is controlled by the substitution of Ge for Si.
The ARPES results reveal that the hole-like FS of Ce 4$f$-electrons become small as the hybridization strength increases.
This indicates that the contribution of Ce 4$f$-electrons to FS increases with the hybridization strength.
It is found that such fine variation of Ce 4$f$-FS is not explained by the LDA band calculations, indicating the importance of the strong correlation effects.

\section{Acknowledgement}
One of the authors (H.J.I.) thanks M. Tsunekawa for fruitful discussion.
This work was funded by the Korean Goverment (KRF-2008-313-C00293) and by Grant-in-Aid for Scientific Research (B) (No.18340110) from MEXT of Japan, and was performed for the Nuclear R\&D Programs funded by the Ministry of Science \& Technology of Korea and under the Proposal No.2007A3835 of SPring-8.
This work was also supported by the Basic Science Research Program (NRF-2013R1A1A2009778) and the Leading Foreign Research Institute Recruitment Program (Grant No. 2012K1A4A3053565).


\end{document}